\documentclass{webofc}
\usepackage[varg]{txfonts}   
\newcommand{\BR}{{\cal B}}
\newcommand{\pp}{\pi^+\pi^-}

\newcommand{\kk}{K^+K^-}

\newcommand{\EE}{e^+e^-}

\newcommand{\etap}{\eta^\prime}
\newcommand{\psp}{\psi(2S)}
\newcommand{\psip}{\psi(2S)}

\newcommand{\jpsi}{J/\psi}
\newcommand{\piz}{\pi^0}

\newcommand{\xyz}{\rm XYZ}
\newcommand{\x}{X(3872)}
\newcommand{\y}{Y(4260)}

\newcommand{\zc}{Z_c(3900)}
\newcommand{\zcp}{Z_c(4020)}

\newcommand{\ppjpsi}{\pi^+\pi^-J/\psi}
\newcommand{\hc}{h_c}
\newcommand{\pphc}{\pi^+\pi^-\hc}

\newcommand{\ccb}{c\bar{c}}
\newcommand{\DDb}{D\bar{D}}
\newcommand{\ppb}{p\bar{p}}

\begin{document}
\title{Exotic states in the quarkonium sector}
\subtitle{--- status and perspectives}

\author{\firstname{Chang-Zheng}
\lastname{Yuan}\inst{1,2}\fnsep\thanks{\email{yuancz@ihep.ac.cn}}
}

\institute{
Institute of High Energy Physics, Chinese Academy of Sciences,
19B Yuquan Road, Beijing, China
\and
University of Chinese Academy of Sciences,
19A Yuquan Road, Beijing, China
          }

\abstract{
The discovery of hadronic states beyond the conventional two-quark meson
and three-quark baryon picture in the last two decades is one of the most amazing 
accomplishments in fundamental physics research. Many experiments
contributed to this field despite of the original goals of the design.
We review the experimental progress on the study of the quarkoniumlike states
--- states with at least one heavy quark-antiquark pair and possible light
quarks, also known as $\xyz$ states.
We give a general review and then focus on the new experimental
results on the $\x$ and its bottom-quark partner $X_b$, the $X(3960)$, 
the $Y(4260)$, $Y(4500)$, $Y(4660)$, and $Y(10750)$, and
the charged charmoniumlike $Z_c$ and $Z_{cs}$ states.
The observations suggest that we did observe hadronic molecules 
and we also observed hadronic states with some other quark configurations.
Possible further studies at the existing and future facilities 
are briefly discussed.
}

\maketitle

\section{Introduction}

Hadron spectroscopy is a field of frequent discoveries and surprises,
and the theoretical difficulties in understanding the strong interaction
in the color-confinement regime make the field even more fascinating.
The tremendous data collected by the BaBar, Belle, BESIII, LHCb,
and other experiments and improved theoretical tools developed to analyze the
experimental data result in rapid progress of the 
field~\cite{Chen:2022asf,Brambilla:2019esw,Guo:2017jvc,Chen:2016qju}.

In the conventional quark model, mesons are composed of one quark
and one anti-quark, while baryons are composed of three quarks.
However, many quarkoniumlike states were discovered at two
$B$-factories BaBar and Belle~\cite{PBFB} in the first decade of
the 21st century. Whereas some of these
are good candidates of quarkonium states, such as the $\eta_c(2S)$, 
$\psi_2(3823)$, $h_b(2P)$, many other states have
exotic properties, which may indicate that exotic states, such as
multi-quark state, hadronic molecule, or hybrid, have been
observed~\cite{Chen:2022asf,Brambilla:2019esw,Guo:2017jvc,Chen:2016qju}.

BaBar and Belle experiments finished their data taking in 2008 and
2010, respectively, and the data are still used for various
physics analyses. BESIII~\cite{ijmpa_review} and
LHCb~\cite{lhcb} experiments started data taking
and contributed to the study of the $\xyz$ particles since 2008.
Most of the discoveries of the such states were made at these
four experiments.

\begin{figure*}[htbp]
\centering
  \includegraphics[width=0.95\textwidth]{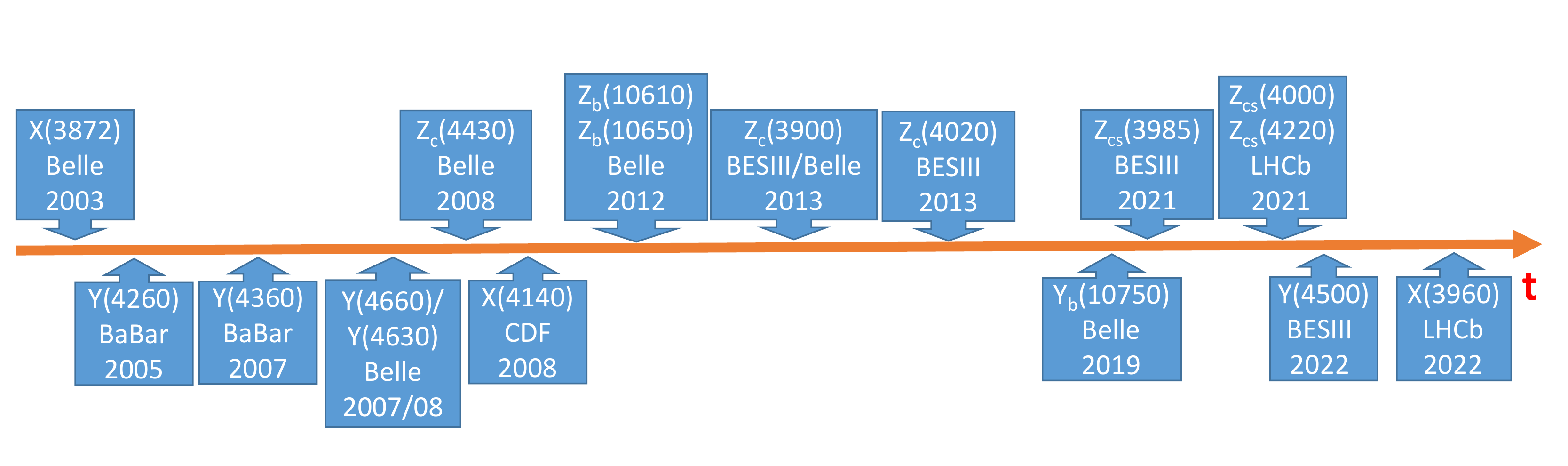}
\caption{Discovery of heavy exotic states from experiments.}\label{fig:XYZ}
\end{figure*}

Figure~\ref{fig:XYZ} shows the history of the discovery of the heavy
exotic states, started from the observation of the $X(3872)$
in 2003~\cite{Bellex}.
In this brief review, We show some recent experimental results on 
these particles, and we focus on those states with exotic properties, 
including the $\x$, $Y(4260)$, $\zc$, and their siblings.

\section{\boldmath The $X$ states}

The $X(3872)$ was observed in 2003 by the Belle
experiment~\cite{Bellex}, and it was confirmed later by CDF~\cite{CDFx} 
and $D0$~\cite{D0x} experiments in $p\bar{p}$ collision. After almost 
20 years' study, we know this state much better than any of the
other similar states. The bottomonium equivalent of the $X(3872)$,
$X_b$, was searched for but not observed, and there are other 
$X$ states observed recently, such as the $X(3960)$ in its decay
into $D_s^+D_s^-$. 

\subsection{\boldmath Resonance parameters of the $X(3872)$}

The mass of the $X(3872)$ has been measured as $3871.65\pm 0.06$~MeV~\cite{pdg},
which is lower than the mass threshold of $\bar{D}^0D^{*0}$, $3871.69\pm 0.11$~MeV,
by $0.04\pm 0.12$~MeV, to be compared with the binding energy of the deuteron
of 2.2~MeV. If the $X(3872)$ is a molecule of $\bar{D}^0D^{*0}$, its size
will be larger than 5~fm, much larger than the size of a typical hadron.

The width measurements are less precise and model dependent since
the $X(3872)$ is very narrow and the mass resolution of the experiments
is usually much larger than the intrinsic width. Fitting the
$\ppjpsi$ invariant mass distribution with a Breit-Wigner function,
LHCb reported a width of about 1~MeV (the mass resolution is 2.4--3.0~MeV);
and the fit with a Flatt\'e function with constraints from other measurements
yields a FWHM of 0.22~MeV which depends strongly on
the $X(3872)\to \bar{D}^0D^{*0}$ coupling~\cite{LHCb_x_width1,LHCb_x_width2}.
Although the statistics are
low at BESIII experiments, the high efficiencies of reconstructing
all the $X(3872)$ decays modes and the very good mass resolution in
the $\bar{D}^0D^{*0}$ mode ($<1$~MeV) make it possible to measure
the line shape of the $X(3872)$ state~\cite{bes3_x_brs}.

\subsection{\boldmath Production of the $X(3872)$}

Production of the $X(3872)$ has been reported in many different
kinds of processes, in $B$ and $B_s$ meson decays, in $\Lambda_b$
baryon decays, in $p\bar{p}$ and $pp$ collisions, and
in $\EE$ annihilation~\cite{BES3x}. Recently, evidence for $X(3872)$
production in $PbPb$ and two-photon collisions, and observation of 
$\EE\to \omega X(3872)$ were reported, whereas no hint of direct 
production of the $X(3872)$ in $\EE$ annihilation was observed. 

CMS experiment reported a $4.2\sigma$ signal of the $X(3872)$ in
$PbPb$ collision at 5.02~TeV~\cite{CMS:2021znk}, and it is interesting to note that
its production rate relative to the $\psip$ is much larger than
in the $pp$ collision at 7 and 8~TeV, although the uncertainty is
large. If this is confirmed, this is a supplemental information
to understand the nature of this state.

Belle experiment searched for the $X(3872)$ in $\gamma\gamma^*$
fusion~\cite{Belle:2020ndp}. Three events are observed in the signal region, 
corresponds to a statistical significance of $3.2\sigma$. Since we know that
the $X(3872)$ has $J^{PC}=1^{++}$, it cannot be produced in two real-photon
collision, the production requires at least one of the photons is virtual.

\begin{figure*}[htbp]
\centering
  \includegraphics[height=4.5cm]{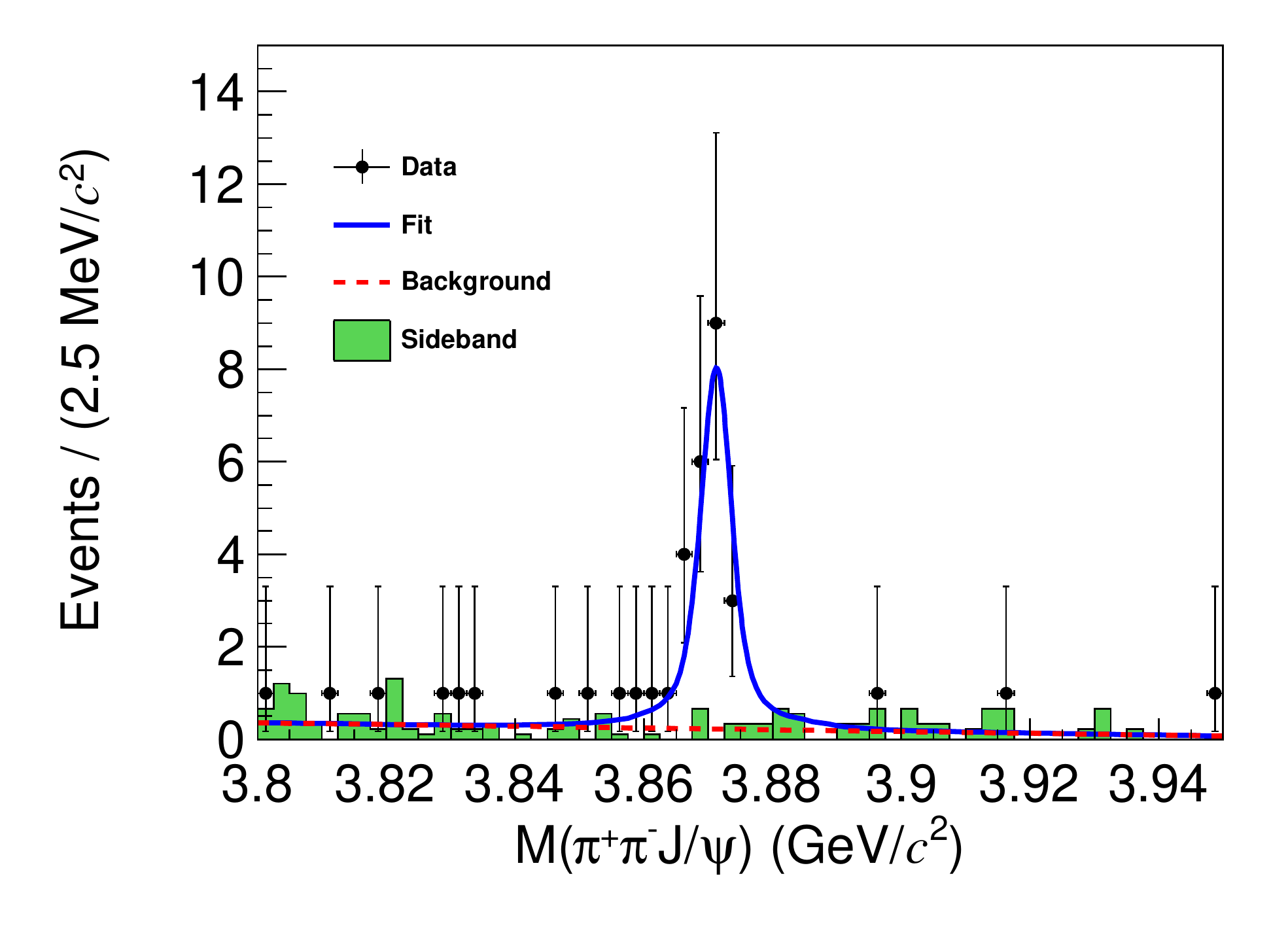}
  \includegraphics[height=4.5cm]{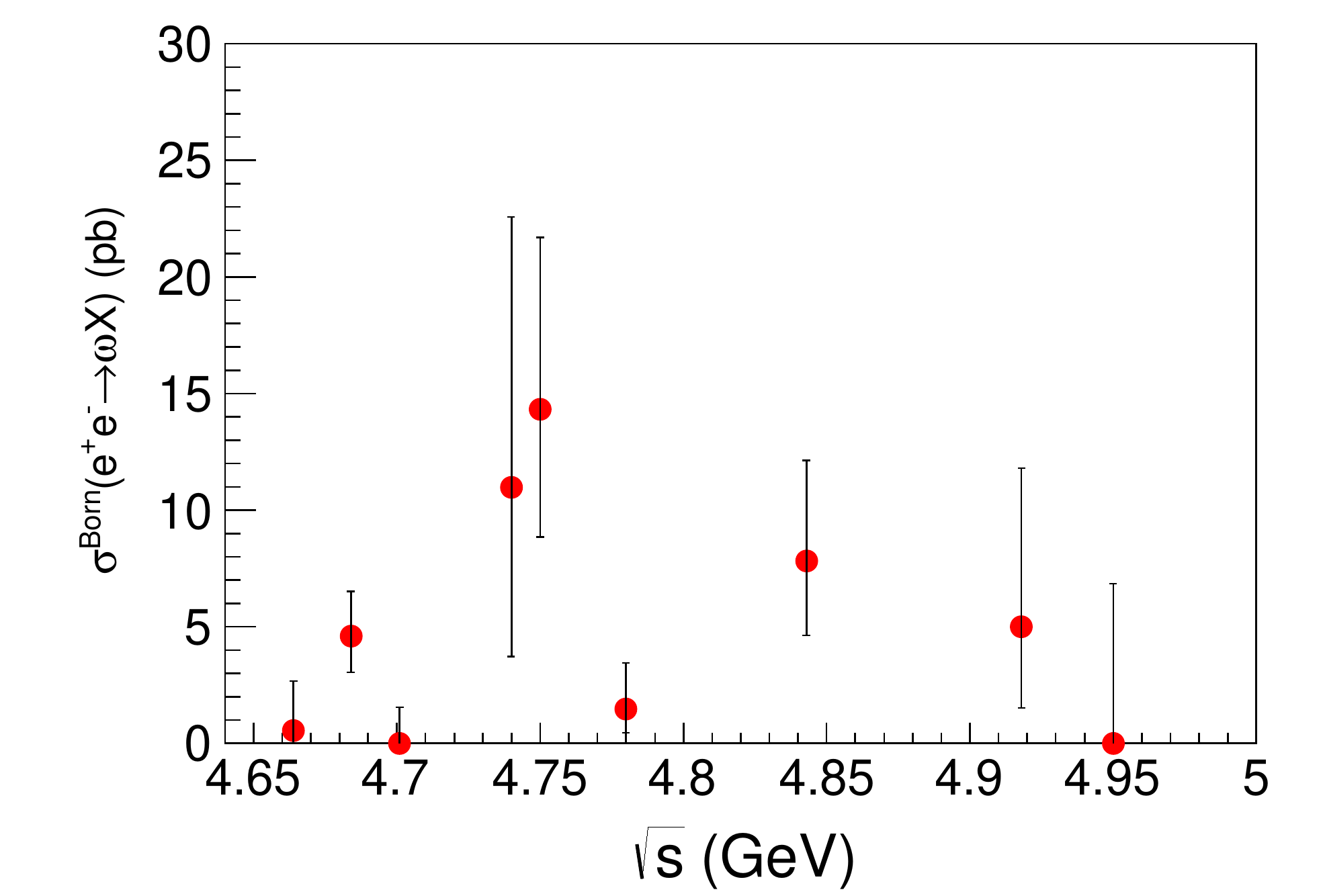}
\caption{The $X(3872)$ signal recoiling against $\omega$ in $\EE$ annihilation 
(left panel) and cross sections of $\EE\to \omega X(3872)$ (right panel) 
from BESIII experiment~\cite{omegaX_lich}.}
\label{fig:omegaX}
\end{figure*}

BESIII experiment reported observation of $\EE\to \omega X(3872)$ with 
4.7~fb$^{-1}$ data at center-of-mass (CM) energies from 4.66 to 4.95~GeV, 
24 $X(3872)$ signal events are observed with a significance of $7.5\sigma$, 
including both the statistical and systematic uncertainties~\cite{omegaX_lich}. 
The $X(3872)$ signal and the cross section as a function of CM energy are 
shown in Fig.~\ref{fig:omegaX}. Although not very conclusive, it seems 
that the $\omega X(3872)$ signal comes from a resonance decay with a mass
of about 4.75~GeV and a peak cross section of around 14~pb.

\begin{figure*}
    \centering
    \includegraphics[height=4.2cm]{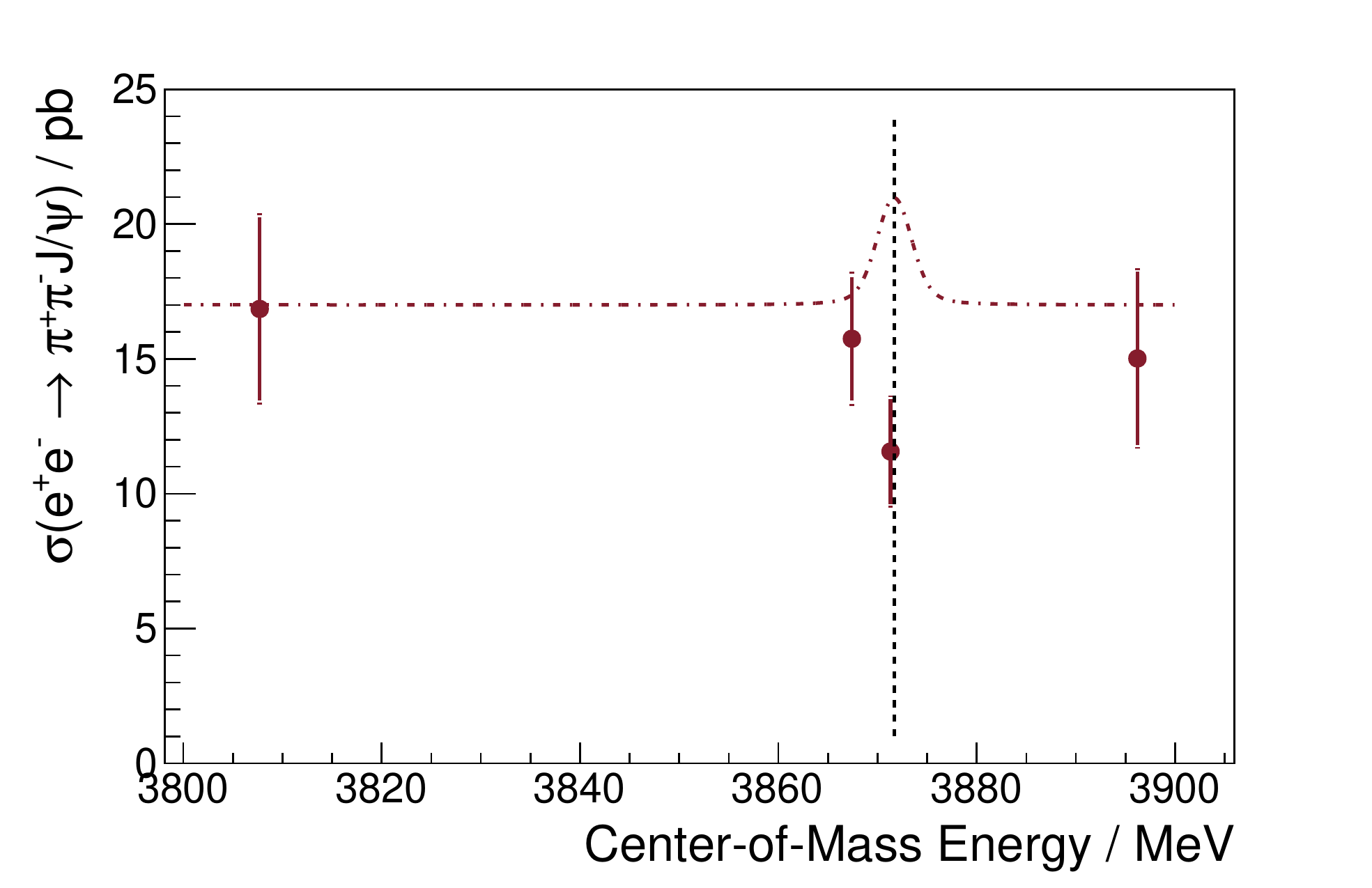}
    \includegraphics[height=4.2cm]{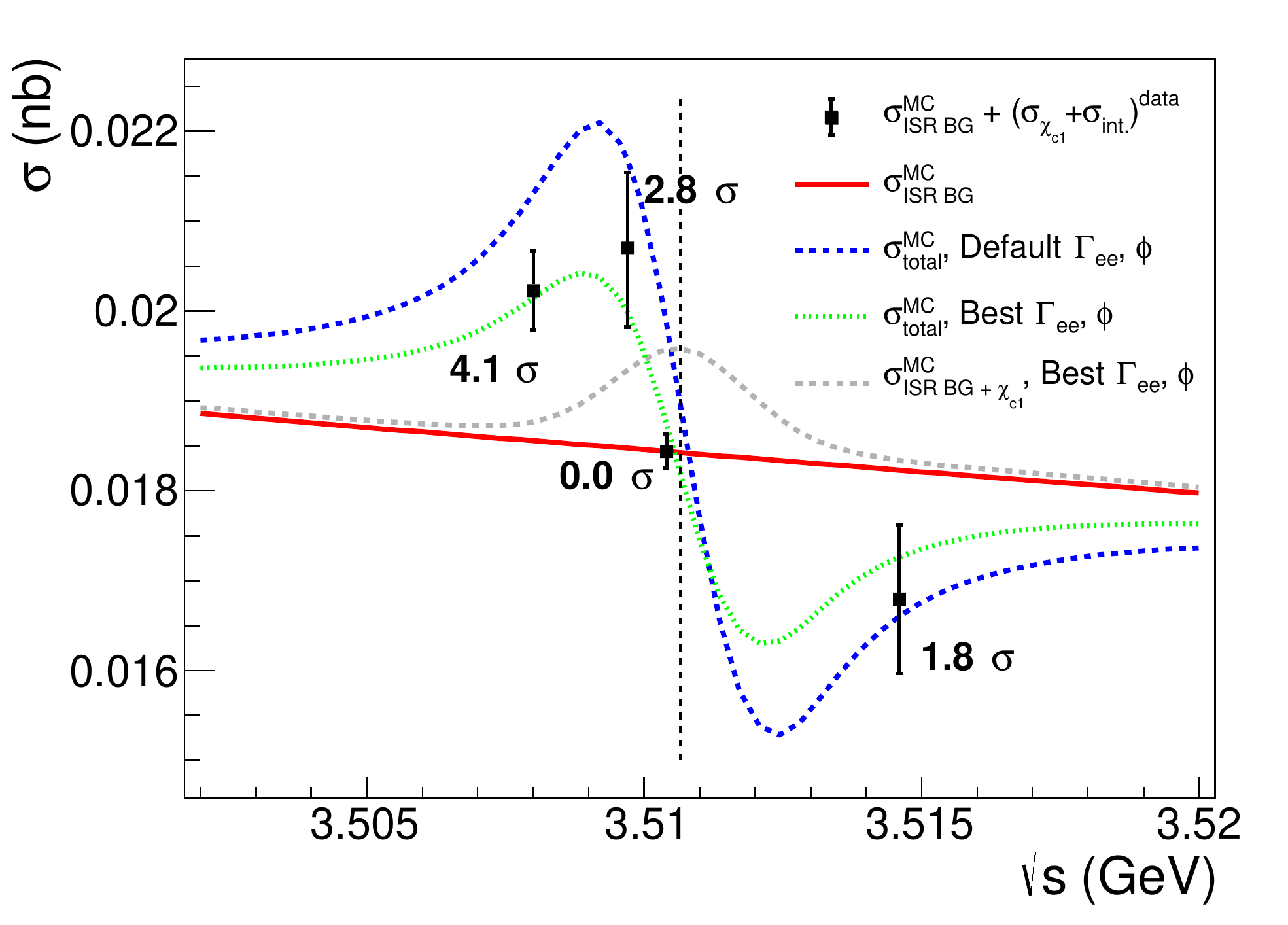}
\caption{Cross sections of $\EE\to \ppjpsi$ in the vicinity of the $X(3872)$ 
(left panel,~\cite{BESIII:2022ner}) and those of $\EE\to \gamma\chi_{c1}$ 
in the vicinity of the $\chi_{c1}$ (right panel,~\cite{BESIII:2022mtl}). }
\label{fig:xs-Xandchi}
\end{figure*}

BESIII experiment searched for $\EE\to X(3872)$ by taking data at a 
CM energy corresponds to exactly the $X(3872)$ mass and at a few energies 
in the vicinity of the $X(3872)$~\cite{BESIII:2022ner} and measure the 
cross sections for the process $\EE\to \ppjpsi$ (see Fig.~\ref{fig:xs-Xandchi}). 
No enhancement of the cross section is observed at the $X(3872)$ peak 
and an upper limit is determined to be
$\Gamma_{ee}\times \BR(X(3872)\to \ppjpsi)<7.5\times 10^{-3}~{\rm eV}$
at the 90\% confidence level, and with the $\BR(X(3872)\to\ppjpsi)$
from PDG~\cite{pdg} as input, an upper limit 
on the electronic width $\Gamma_{ee}$ of $X(3872)$ is
obtained to be $<0.32$~eV at the 90\% confidence level. 
Since the process $\EE\to \chi_{c1}$ has been observed 
(see right panel of Fig.~\ref{fig:xs-Xandchi}) by the BESIII 
experiment~\cite{BESIII:2022mtl}, it is only a matter of sensitivity of the 
experiment to observe $\EE\to X(3872)$ since both $\chi_{c1}$ and $X(3872)$ 
have the same quantum numbers ($J^{PC}=1^{--}$) and the expected
electronic width of the $X(3872)$ is at the same level as the $\chi_{c1}$.
Although not significant, the lower cross section at the $X(3872)$ peak
than in the nearby energies may indicate interference between 
$\EE\to X(3872)$ and non-resonant $\EE\to \rho\jpsi$~\cite{Davier:2006fu,KTChao} 
amplitudes.

\subsection{\boldmath Decays of the $X(3872)$}

The total production rate of the $X(3872)$ in $B$ decays was measured
by reconstructing a $B^-$ and a charged kaon from $B^+$ decays and
checking the recoiling mass of the $B^-K^+$ system. BaBar observed a small peak,
corresponding to a $3.0\sigma$ significance at the $X(3872)$ signal region,
and measured the branching fraction of $B^+\to K^+X(3872)$ as
$(2.1\pm 0.6\pm 0.3)\times 10^{-4}$~\cite{BaBar:2019hzd}. Belle did the same analysis,
but the signal is less significant and the resulting branching fraction
is $(1.2\pm 1.1\pm 0.1)\times 10^{-4}$ and the signal significance
is $1.1\sigma$~\cite{Belle:2017psv}. Although the signals are not very significant,
we know this process must exist because this state has been observed in its
many exclusive decays. One can use these measurements combined with
other information, such as the product branching fractions and
the ratio of the branching fractions, to determine the decay branching
fractions of the $X(3872)$, including its decays to open charm final states,
hadronic transitions, and radiative transitions.
There could be a small branching fraction to light hadrons, but
no experiment has observed any of them.

The authors of Ref.~\cite{Li:2019kpj} did a global fit to the currently
available experimental measurements of the product branching fractions and
the ratios of the branching fractions. It is found that the branching fraction
of open charm decay is around 50\% and that of each hadronic transition
is at a few per cent level, there is still around one-third of the
$X(3872)$ decays unknown. A few searches for the new decay modes 
of the $X(3872)$ were reported recently, and more will be searched for in 
the future experiments like BESIII and Belle II.

BESIII searched for $X(3872)\to \piz\chi_{c0}$ and $\pp\chi_{c0}$ 
with 9.9~fb$^{-1}$ data at CM energies between 4.15 and 
4.30~GeV~\cite{BESIII:2022kow}. No signals are observed and the
upper limits at the 90\% C.L. are determined as 
$\frac{\BR(X(3872)\to \piz\chi_{c0})}{\BR(X(3872)\to \pp\jpsi)}<3.6$,
$\frac{\BR(X(3872)\to \pp\chi_{c0})}{\BR(X(3872)\to \pp\jpsi)}<0.56$, and
$\frac{\BR(X(3872)\to \piz\piz\chi_{c0})}{\BR(X(3872)\to \pp\jpsi)}<1.7$.

Belle reported a search for $X(3872)\to \pp\piz$ in 
$B^{\pm}\to K^{\pm}X(3872)$ and $B^{0}\to K_S^0 X(3872)$ 
decays~\cite{Belle:2022puc}. 
No signal is observed and the 90\% credible upper limits are set 
for two different models of the decay processes: if the decay 
products are distributed uniformly in phase space,
$\BR(X(3872)\to \pp\piz) < 1.3\%$; if $M(\pp)$ is concentrated 
near the mass of the $\DDb$ pair in the process 
$X(3872)\to D^0\bar{D}^{*0}+c.c.\to\DDb\piz\to \pp\piz$,
$\BR(X(3872)\to \pp\piz) < 1.2\times 10^{-3}$.

LHCb reported a detailed analysis of the $m_{\pp}$ distribution of
$X(3872)\to \pp\jpsi$, contributions from $\omega\to \pp$ and
its interference with $\rho^0\to \pp$ are observed with high 
significance and the isospin-violating effect in $X(3872)$ decays
is measured with improved precision~\cite{LHCb:2022bly}.

One still very confusing decay mode is $X(3872)\to \gamma \psp$.
There have been four different measurements. The BaBar experiment
claimed $3.5\sigma$ evidence of this mode and a production rate
relative to  $X(3872)\to \gamma \jpsi$ is $3.4\pm 1.4$~\cite{R_in_babar},
but Belle failed to find significant signal and the ratio was
measured to be less than $2.1$ at the 90\% C.L.~\cite{R_in_belle}.
Three years later, LHCb did the same analysis and the found
a $4.4\sigma$ signal with a ratio of $2.46\pm 0.81$~\cite{R_in_lhcb}, 
but a recent BESIII measurement found no signal and a much stringent
upper limit of the ratio is determined to be 0.59 at the 90\% 
C.L.~\cite{bes3_x_brs}. We have four experiments here, 
two claimed evidence and the other two observed nothing. So it is still 
not clear whether this channel, $X(3872)\to \gamma \psp$, exists, 
or if it exists, how small the branching fraction is.

\subsection{\boldmath $X_b$, bottomonium equivalent of the $X(3872)$}

\begin{figure*}[htbp]
\centering 
\includegraphics[width=0.8\textwidth]{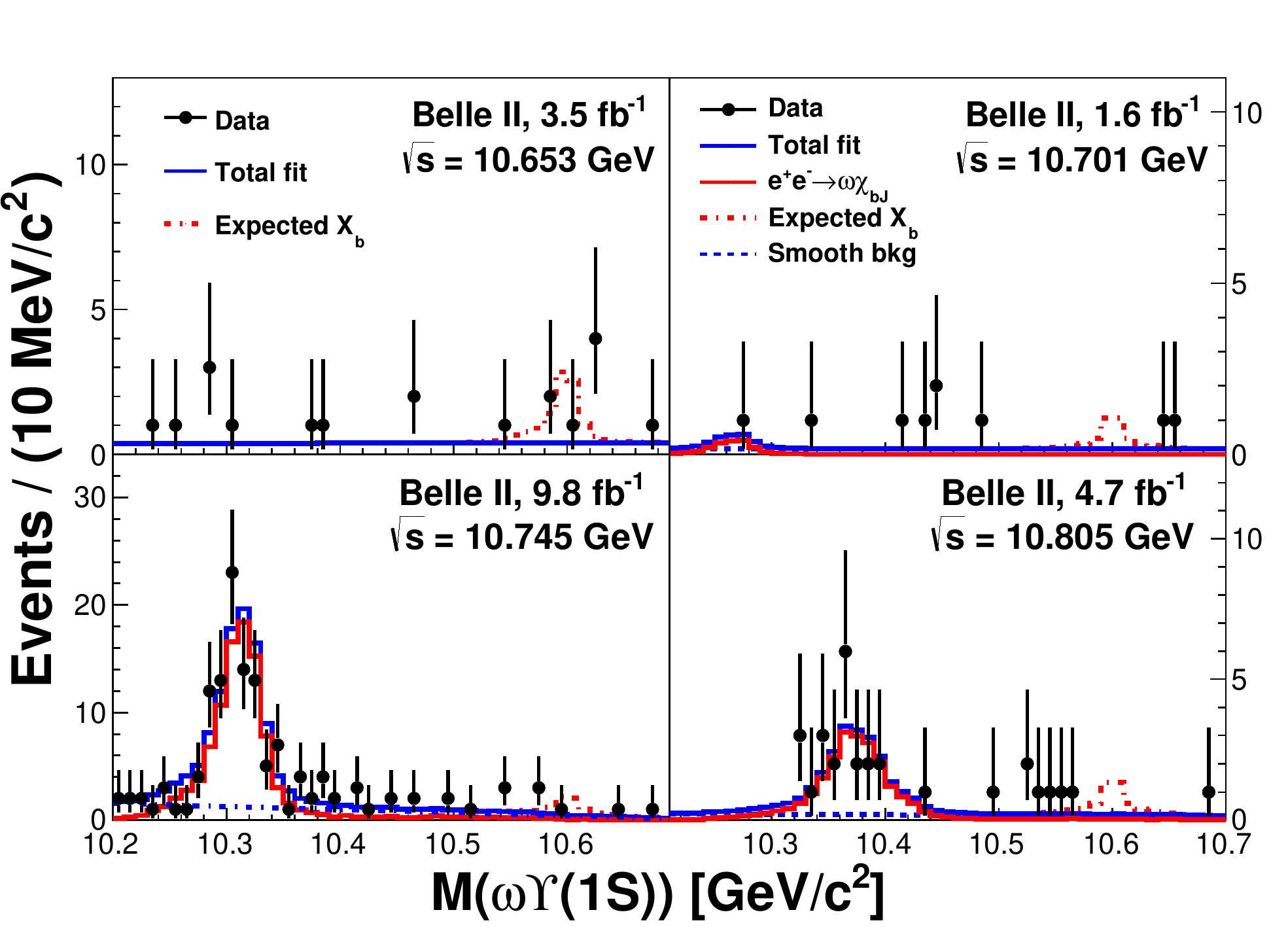}
\caption{Invariant mass distributions of $\omega\Upsilon(1S)$ from Belle
and Belle II data at $\sqrt{s}$ = 10.653, 10.701, 10.745, and 
10.805~GeV~\cite{Belle-II:2022xdi}.  
The red dash-dotted histograms are from simulated events $\EE\to
\gamma X_b (\to \omega\Upsilon(1S))$ with the $X_b$ mass fixed at
10.6~GeV and yields fixed at the upper limit values.}
\label{Xb-belle2}
\end{figure*}

Belle II experiment collected data at CM energies 
10.701, 10.745, and 10.805~GeV and combined with the Belle data
at 10.653~GeV to search for the
bottomonium equivalent of the $X(3872)$ state, $X_b$, decaying into
$\omega\Upsilon(1S)$~\cite{Belle-II:2022xdi}. No significant signal 
is observed for $X_b$ masses between 10.45 and 10.65~GeV
(Fig.~\ref{Xb-belle2}).

\subsection{\boldmath Observation of the $X(3960)$}

LHCb did an amplitude analysis of the $B^+\to D_s^+ D_s^- K^+$ decay 
using proton-proton collision data collected at CM energies of 7, 8 and 
13~TeV~\cite{LHCb:2022vsv}. About 360 signal events are identified,
and a near-threshold peaking structure, $X(3960)$, is observed 
in the $D_s^+ D_s^-$ invariant-mass spectrum with significance 
greater than $12\sigma$. The mass, width and the quantum numbers 
of the structure are measured to be $3956\pm 5\pm 10$~MeV, 
$43\pm 13\pm 8$~MeV, and $J^{PC}=0^{++}$, respectively. Further 
investigation is needed to understand the nature of this state. 

Such a state could be produced in the radiative transitions of
the $Y$ and excited $\psi$ states such as the $\psi(4040)$, 
$\psi(4160)$, $Y(4260)$, $\psi(4415)$. It can be searched for 
with the large data samples at BESIII experiment.

\section{\boldmath The $Y$ states}

\subsection{\boldmath The charmoniumlike $Y$ states}

The $Y$ states were discovered in the initial state radiation (ISR) in
the $B$-factory experiments, and they have $J^{PC}=1^{--}$.
So these state can also be produced directly in $\EE$ annihilation
experiment like BESIII. In this case, much larger statistics
are achieved and these states, the $Y(4260)$, $Y(4360)$, $Y(4660)$,
and so on, are measured with improved precision.

The $Y(4260)$ was observed in 2005 by BaBar experiment~\cite{babar_y4260} 
and the most precise measurement is from the BESIII 
experiment~\cite{BESIII:2022qal} (an update of the analysis reported
in Ref.~\cite{BESIII:2016bnd}). By doing a high
luminosity energy scan in the vicinity of the $Y(4260)$,
BESIII found the peak of the $Y(4260)$ is much lower (so now named
the $Y(4230)$) than that from
previous measurements and the width is narrow, and there is a high
mass shoulder with a mass of 4.3~GeV if fitted with a BW
function. Since then, more new decay modes of the $Y(4230)$ were
observed including $\pphc$, $\omega\chi_{c0}$, 
$\pi \bar{D} D^*+c.c.$, $K^+K^-\jpsi$, and so on, and
no significant $Y(4230)$ was observed in $\EE\to \pp D^+D^-$
process~\cite{BESIII:2022quc} from a recent BESIII measurement.

A global fit~\cite{Gao:2017sqa} to four modes ($\pp\jpsi$, $\pphc$, 
$\omega\chi_{c0}$, and $\pi \bar{D} D^*+c.c.$) was performed, and 
the mass of the $Y(4230)$ as $4230\pm 6$~MeV and the width of 
$56\pm 8$~MeV are determined. It is interesting to point
out that the mass of this resonance is quite close to the
threshold of threshold of $D_s^{*+}D_s^{*-}$ which is $4224$~MeV.
Since there are more $Y(4230)$ decay modes observed
($\pp\psp$, $\eta_c\pp\piz$, $K^+K^-\jpsi$, and so on),
this combined fit can be updated with more channels.

\begin{figure*}
\begin{center}
\includegraphics[height=5.cm]{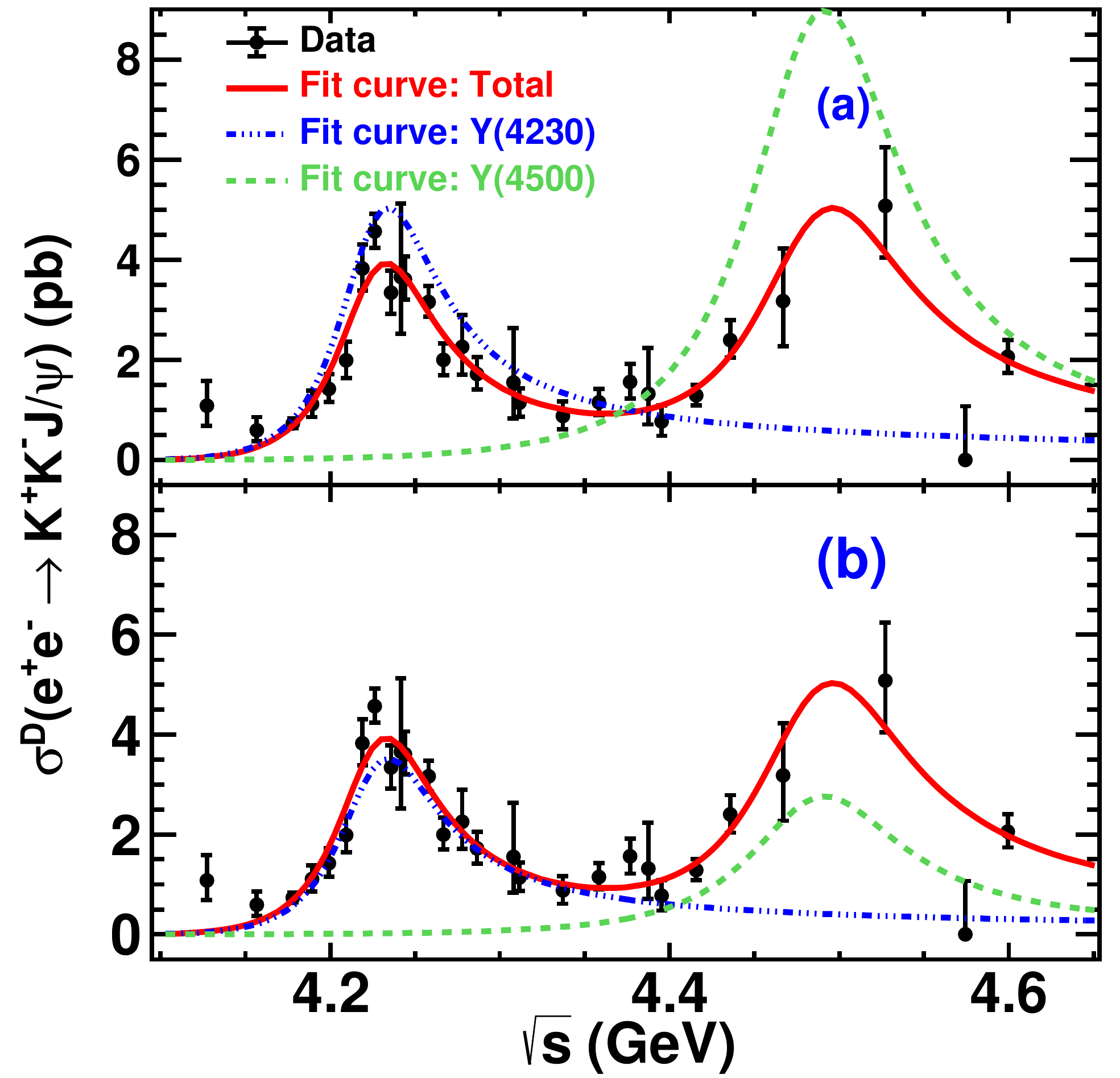}
\includegraphics[height=5.cm]{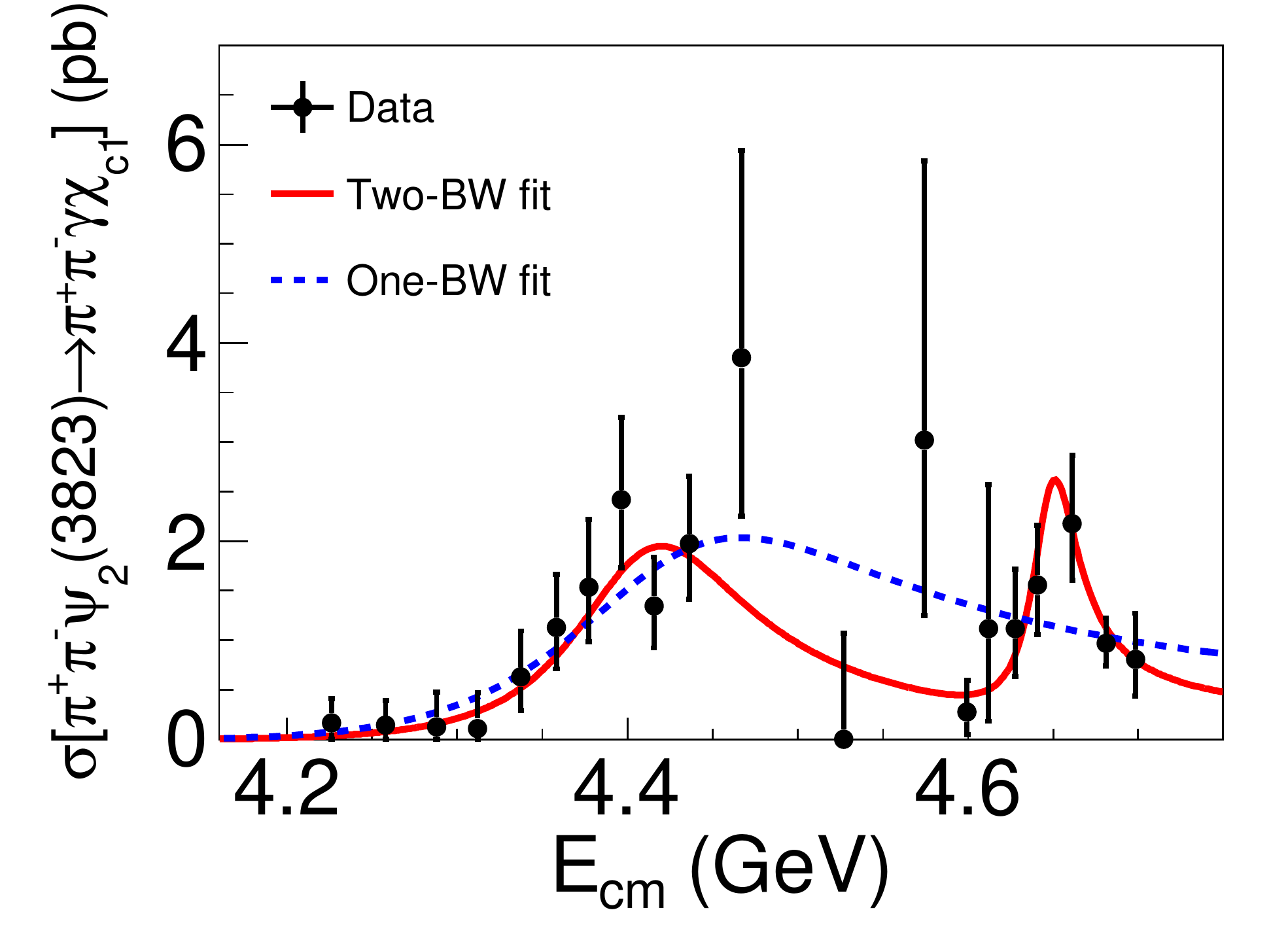}
\caption{Fits to $\sigma(e^+e^- \to K^+K^-\jpsi)$ 
(Left panel,~\cite{BESIII:2022joj}) and $\sigma[\EE\to\pp \psi_2(3823)]\times
\BR[\psi_2(3823)\to\gamma\chi_{c1}]$ (right panel,~\cite{BESIII:2022yga}). }
\label{xs-y4500-y4660}
\end{center}
\end{figure*}

Recently, the cross sections of $\EE\to K^+K^-\jpsi$ at
CM energies from 4.1 to 4.6~GeV are measured 
at the BESIII~\cite{BESIII:2022joj}. 
Two resonant structures are observed in the line shape of 
the cross sections (see Fig.~\ref{xs-y4500-y4660}). 
The mass and width of the first structure 
are measured to be ($4225.3\pm2.3\pm21.5$)~MeV and 
($72.9\pm6.1\pm30.8$)~MeV, respectively. 
They are consistent with those of the established $Y(4230)$. 
The second structure is observed for the first time
with a statistical significance greater than $8\sigma$, denoted as
the $Y(4500)$. Its mass and width are determined to be
($4484.7\pm13.3\pm24.1$)~MeV and ($111.1\pm30.1\pm15.2$) MeV,
respectively. The product of the electronic
partial width with the decay branching fraction 
$\Gamma(Y(4230)\to e^+ e^-) \BR(Y(4230) \to K^+ K^-
\jpsi)$ is found to be $1.35\pm0.14\pm0.07$~eV or 
$0.41\pm0.08\pm0.13$~eV. This state is consistent with 
a vector charmonium state in the 5S-4D mixing scheme~\cite{Wang:2019mhs}, 
a heavy-antiheavy hadronic molecule~\cite{Dong:2021juy}, or 
a $(cs\bar{c}\bar{s})$ tetraquark state~\cite{Chiu:2005ey}.

For the state at 4.66~GeV, it was observed in $\EE\to \pp\psp$ 
by Belle~\cite{belle_y4660}, and confirmed by BaBar~\cite{BaBar:2012hpr}. 
The peak position is at around 4.66~GeV,
thus it is called the $Y(4660)$. There is another state observed
in $\EE\to \Lambda_c \bar{\Lambda}_c$
by the Belle experiment~\cite{Belle:2008xmh}, but the peak is at
around 4.63~GeV, although the error is large. It is not clear
whether these two states are the same or whether there are two states
in this energy region.

BESIII data on $\EE\to \pp\psp$ mode from 4.0 to 4.7~GeV confirmed
the Belle and BaBar observations with much improved precision~\cite{BESIII:2021njb},
BESIII has data now covering from threshold to 4.95~GeV,
comparable precision as at 4.6~GeV is expected at high energies, so we expect
better measurement of the $Y(4660)$ state from BESIII soon.

BESIII measured the product of the $\EE\to \pi^+\pi^-\psi_2(3823)$ 
cross section and the branching fraction $\BR[\psi_2(3823)\to
\gamma\chi_{c1}]$ at CM energies from $4.23$ to 
$4.70$~GeV~\cite{BESIII:2022yga}. For the first time, resonance 
structure is observed in the cross section line shape of $e^+e^-\to
\pi^+\pi^-\psi_2(3823)$ with significances exceeding $5\sigma$ 
(see Fig.~\ref{xs-y4500-y4660}). A fit to data with 
two coherent Breit-Wigner resonances modeling the
$\sqrt{s}$-dependent cross section yields $M(R_1)=4406.9\pm
17.2\pm 4.5$~MeV, $\Gamma(R_1)=128.1\pm 37.2\pm 2.3$~MeV,
and $M(R_2)=4647.9\pm 8.6\pm 0.8$~MeV, $\Gamma(R_2)=33.1\pm
18.6\pm 4.1$~MeV. Though weakly disfavored by the data, a single
resonance with $M(R)=4417.5\pm26.2\pm3.5$~MeV,
$\Gamma(R)=245\pm48\pm13$~MeV is also possible to interpret data.
Within current uncertainties, the parameters of structures in the
two resonances interpretation are similar to the $Y(4360)$ and 
$Y(4660)$ states reported in $\pp\psip$~\cite{belle_y4660,BaBar:2012hpr}.
Assuming the observed structures correspond to these resonances,
this will be the second decay channel of the $Y(4660)$
state after more than 15 years of its discovery. 

Belle reported measurements of two open-charm final states.
There is a very beautiful peak observed at around 4.63~GeV
in $D_s^+D_{s1}(2536)^-+c.c.$ mode and the signal significance
is $5.9\sigma$~\cite{Belle:2019qoi}. The signal in $D_s^+D_{s2}(2573)^-+c.c.$ 
mode is not so significant, is only $3.4\sigma$~\cite{Belle:2020wtd}.

If we put all these information together, we can find that
the peak position is about 4.65~GeV in $\pp\psp$ and $\pi^+\pi^-\psi_2(3823)$
modes, and that in open charm baryon and meson pair final states is
below 4.65~GeV, There are differences from different final states.
We need more measurements to really understand the structures
in this mass region.

\subsection{\boldmath The bottomoniumlike $Y(10750)$}

A Belle study of $\EE\to \pp\Upsilon(nS)$ ($n=1,~2,~3$) revealed the 
existence of a new vector bottomoniumlike state, the $Y(10750)$, 
with a mass of $(10752.7\pm 5.9^{+0.7}_{-1.1})$~MeV and width 
$(35.5^{+17.6}_{-11.3}{}^{+3.9}_{-3.3})$~MeV~\cite{Y10750}.
However, this state is at exactly the position of a dip in 
the total cross section of $\EE\to b \bar{b}$~\cite{Dong:2020tdw}. 
This indicates that the dip is very likely produced by the interference 
between a resonance and a smooth background amplitudes.

\begin{figure*}[htbp]
\centering 
\includegraphics[width=0.90\textwidth]{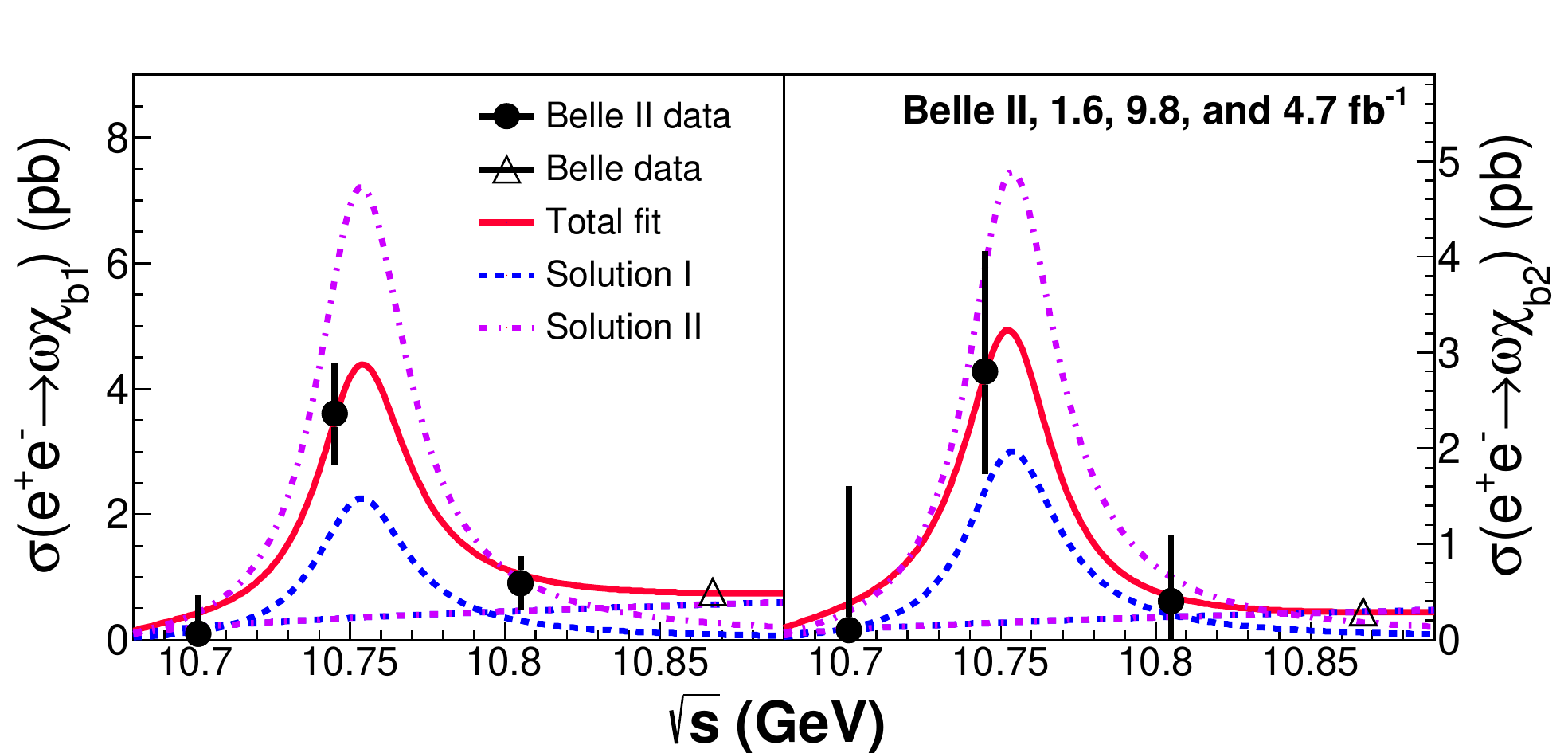}
\caption{Energy dependence of the cross sections for 
$e^+e^- \to \omega\chi_{b1}$ (left panel) and 
$e^+e^- \to \omega\chi_{b2}$ (right panel)~\cite{Belle-II:2022xdi}. 
Curves show the fit results and various components of the fit function.}
\label{soluchib}
\end{figure*}

To understand the $Y(10750)$ better, Belle II experiment accumulated
data at CM energies 10.701, 10.745, and 10.805~GeV, 
corresponding to 1.6, 9.8, and 4.7~fb$^{-1}$ of integrated luminosity, 
respectively. Belle II reported the first observation~\cite{Belle-II:2022xdi} 
of $\EE\to \omega\chi_{bJ}(1P)$ ($J=1,~2$) signals and by combining the 
Belle II data with Belle results at $\sqrt{s}$ = 10.867~GeV, 
Belle II found that the energy dependencies of
the cross sections for $\EE\to \omega\chi_{b1,b2}(1P)$ are 
consistent with the shape of the $\Upsilon(10750)$ state
(shown in Fig.~\ref{soluchib}).
By fitting the energy dependence of the cross sections for
$\EE\to \omega\chi_{b1}$ and $\omega\chi_{b2}$, one obtains
$\Gamma_{ee}\BR(\Upsilon(10750)\to\omega\chi_{b1}$ 
and $\omega\chi_{b2})$ in the range 0.20--2.9 and 0.05--2.0~eV,
respectively. 

\section{Charged quarkonium states}

These include the $Z_c$, $Z_b$, and also the $Z_{cs}$ states.
Since these states decay into final states with one pair of
heavy quarks and charged, there must be at least four quarks
in their configurations.

The $Z_c(3900)$ discovered by BESIII~\cite{zc3900} and
Belle~\cite{Belle_zc} is quite close to the
$\bar{D}D^*$ threshold, and the $Z_c(4020)$ discovered by BESIII is
quite close to the $\bar{D}^*D^*$ threshold~\cite{BESIII:2013ouc},
while the $Z_c(4430)$ discovered by the Belle~\cite{Belle_zc4430}
is not quite close to any of the open charm
threshold. In the bottom sector, The $Z_b(10610)$ and $Z_b(10650)$
discovered by Belle~\cite{Belle:2011aa} are close to the $\bar{B}B^*$
and $\bar{B}^*B^*$ thresholds, respectively.

These states have been observed for some time. Recent studies try to
search for states with one of the four quarks replaced by a different
quark, for example, the $Z_{cs}$ states with quark content 
$\ccb u \bar{s}$ or $\ccb d \bar{s}$.
There are three different measurements, one from Belle in
$\EE\to\kk\jpsi$~\cite{Belle:2014fgf}, another from BESIII in 
$\EE\to K(DD_s^*+D^*D_s)$~\cite{BESIII:2020qkh,BESIII:2022qzr},
and the third from LHCb in $B^+\to\jpsi K^+ \phi$~\cite{LHCb:2021uow}.

\begin{figure*}[htbp]
\centering 
\includegraphics[width=0.95\textwidth]{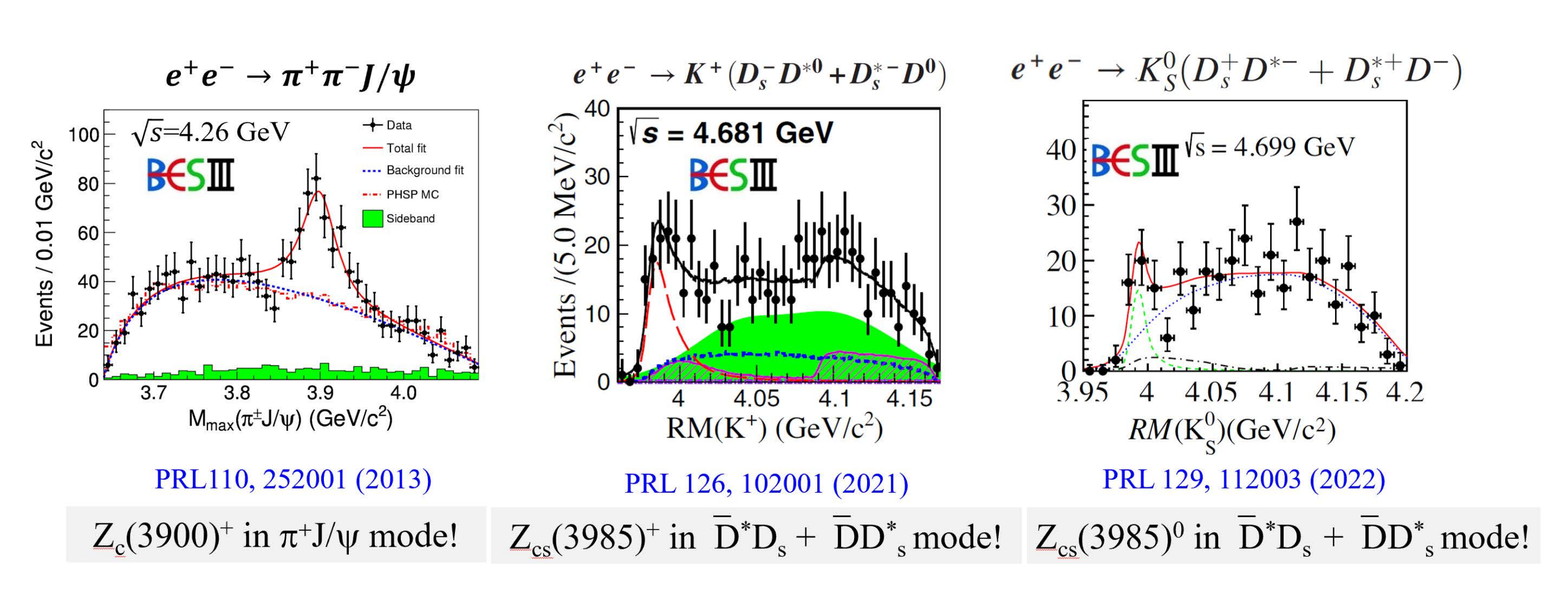}
\caption{The $\zc$ (left panel,~\cite{zc3900}), $Z_{cs}(3985)^+$ 
(middle panel,~\cite{BESIII:2020qkh}), and 
$Z_{cs}(3985)^0$ (right panel,~\cite{BESIII:2022qzr}) signals 
observed in BESIII experiment.}
\label{Zc-bes3}
\end{figure*}

No significant signal was observed in Belle data of ISR production
of $\EE\to\kk\jpsi$~\cite{Belle:2014fgf}. BESIII announced observation of a
near-threshold structure $Z_{cs}(3985)$ in the $K^+$ recoil-mass spectrum
in $\EE\to K^+(D^-_sD^{*0}+D^{*-}_sD^{0})$~\cite{BESIII:2020qkh} with a mass of
3983~MeV and a width of about 10~MeV and evidence for its neutral partner
in the $K^0_S$ recoil-mass spectrum in 
$\EE\to K^0_S(D^-_sD^{*+}+D^{*-}_sD^{+})$~\cite{BESIII:2022qzr}
(see Fig.~\ref{Zc-bes3}). 
LHCb reported two resonances decaying into $K^\pm\jpsi$, 
the $Z_{cs}(4000)$ with a mass of 4003~MeV and a width of about 131~MeV, 
and the $Z_{cs}(4220)$ with a mass of 4216~MeV and a width of about
233~MeV~\cite{LHCb:2021uow}.

\begin{figure*}[htbp]
\centering 
\includegraphics[width=0.85\textwidth]{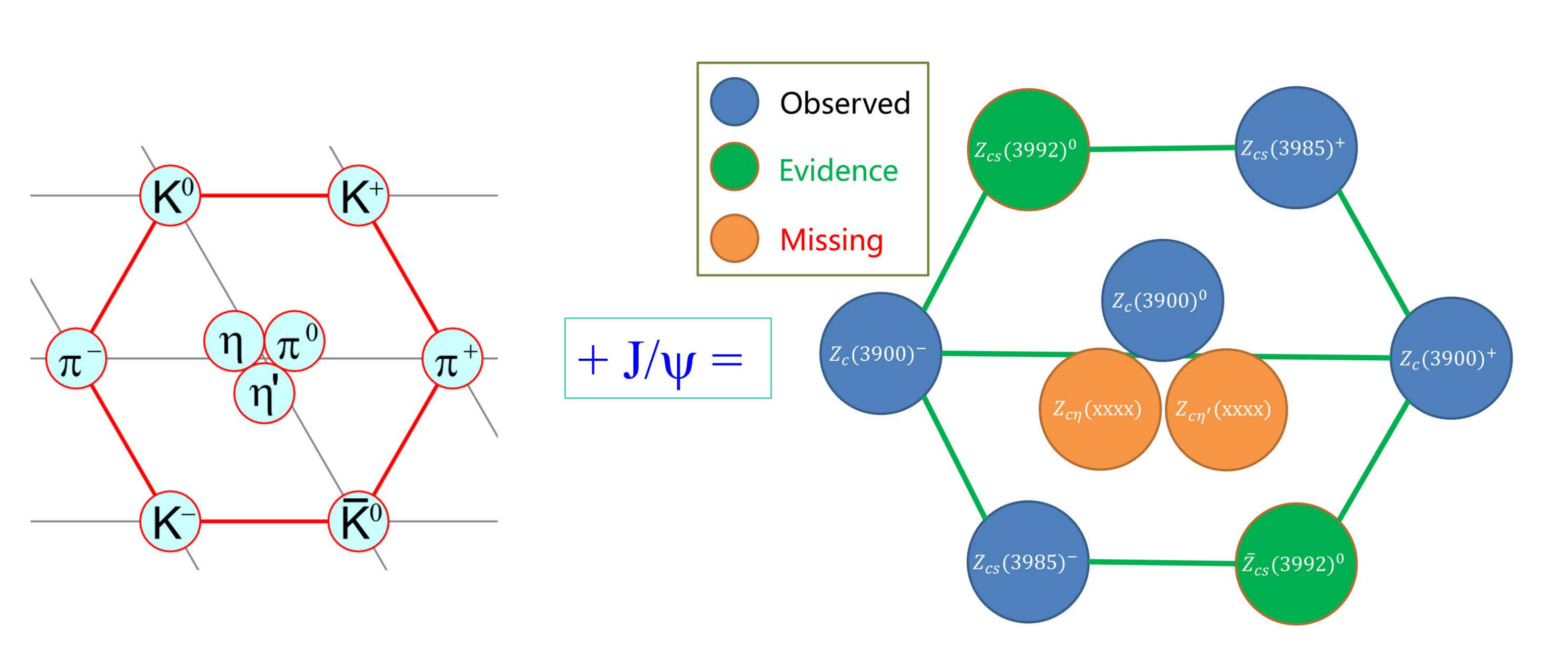}
\caption{The similarity of the pseudoscalar nonet and the $J^P=1^+$ $Z_c$ nonet.}
\label{Zc-nonet}
\end{figure*}

The widths of the $Z_{cs}(3985)$ and $Z_{cs}(4000)$ are quite different,
so they could not be the same state. Maybe one of them is the strange
partner of the $Z_c(3900)$ with the $d$ quark replaced with an $s$ quark. 
These may suggest the existence of a $J^P=1^+$ nonet similar to 
the lowest lying pseudoscalar nonet (see Fig.~\ref{Zc-nonet}), and the
states correspond to $\eta$ and $\etap$ need to be further searched for.

\section{Perspectives}

Although Belle and BaBar have ended their data taking for more than 
10 years, there are still analyses ongoing with the existing data samples.

BESIII has produced a considerable amount of information about the
$\xyz$ and the conventional charmonium states~\cite{xyz-NSR}. 
In addition, there are data that are still being analyzed and more 
data that will be accumulated at other CM energies~\cite{fop,bes3_white}. 
Analyses with these additional data samples will provide improved
understanding of the $\xyz$ states, especially the $\x$, $\y$,
$\zc$, and $\zcp$. The maximum CM energy accessible at BEPCII
was upgraded from 4.6 to 5.0~GeV in 2019, and 5.6~fb$^{-1}$
of data were accumulated in the 2019-20 and 2020-21 running
periods, with more data planned for the future. This enables a
full coverage of the $Y(4660)$~\cite{belle_y4660} resonance and a
search for possible higher mass vector mesons and states with
other quantum numbers, as well as improved measurements of their
properties. A further upgrade of the accelerator will enable
an energy coverage up to 5.6~GeV and with a factor of 3 improvement
of the luminosity at above 4.7~GeV~\cite{bepc2u}. This will enable a study
of the CM energy region 4.7 to 5.6~GeV that was not well investigated
due to lack of data~\cite{BESIII:2022mxl}. It will take about 3 years
to prepare the upgraded components and half a year for installation
starting from summer 2024, and commissioning is planed in early 2025. 

At the same time, the B-factory experiments will supply substantial
information on these states and possibly discover 
more~\cite{LHCb:2022ine,belle2}. At the LHCb, in addition to the 9~fb$^{-1}$ 
of data at 7, 8, and 13~TeV that have been used for most of their published
analyses, 50~fb$^{-1}$ more data are being accumulated in run3 which 
was started in summer 2022 and at 13.6~TeV~\cite{LHCb:2022ine}. The huge 
statistics at LHCb and very low background after tagging the long lifetime 
b-hadrons make many searches and precision studies possible. 
The study of final states with photons and $\piz$ will be 
very challenging at LHCb, an alternative way of photon detection
of using gamma-conversion will help but with a considerable
drop of efficiency.

Belle II~\cite{belle2} has collected 424~fb$^{-1}$ of data by
mid-2022, and will accumulate 50~ab$^{-1}$ data at the
$\Upsilon(4S)$ peak by the end of 2035~\cite{Belle-II:2022cgf}. 
These data samples can be
used to study the $\xyz$ and charmonium states in many different
ways~\cite{PBFB}, among which ISR can produce events in the same
energy range covered by the BESIII. A 50~ab$^{-1}$ Belle II data
sample will correspond to about 250~fb$^{-1}$ of data 
for $\EE$ collision energy between 4 and 5~GeV. 
Similar statistics will be available for modes like
$\EE\to \ppjpsi$ at Belle II and BESIII (after considering the
fact that Belle II has lower efficiency). Belle II has the
advantage that data at different energies will be accumulated at
the same time, making the analysis much simpler than at BESIII.
Belle II is unique in studying the bottomoniumlike states by
doing energy scan above the $\Upsilon(4S)$ peak up to about 11~GeV,
the maximum energy SuperKEKB can reach.

The PANDA Experiment at the Facility for Antiproton and Ion Research (FAIR) 
is under construction and may start commissioning of the experiment
in 2027~\cite{PANDA:2021ozp}. It will be able to study charmoniumlike 
exotic states via $\ppb$ annihilation. The momentum range of the 
antiproton beam is 1--15~GeV and the peak luminosity is 
$2\times 10^{31}~{\rm cm}^{-2}s^{-1}$ (Phase 1+2) and 
$2\times 10^{32}~{\rm cm}^{-2}s^{-1}$ (Phase 3). The extremely good
precision of the beam energy measurement will enable very precise
line shape scan of the narrow resonances like $X(3872)$~\cite{PANDA:johan}.

There are two super $\tau$-charm factories being proposed, the STCF 
in China~\cite{HIEPA} and the SCT in Russia~\cite{SCT_charm2018}.
Both machines would run at CM energies of up to 5~GeV or higher
with a peak luminosity of $10^{35}$~cm$^{-2}$s$^{-1}$ which is a
factor of 100 improvement over the BEPCII. These would enable
systematic studies of the charmoniumlike $\xyz$ states with
unprecedented precision.

\section{Summary}

If we summarize these quarkoniumlike hadrons, we find that some of 
them are quite close to the thresholds of two heavy flavor mesons,
like the $X(3872)$ ($\bar{D}^0D^{*0}$),
$Y(4220)$ ($D_s^{*+}D_s^{*-}$ or $\bar{D}D_1)$,
$\zc^+$ ($\bar{D}^0D^{*+}$),
$\zcp^+$ ($\bar{D}^{*0}D^{*+}$),
$Z_{cs}(3985)^+$ or $Z_{cs}(4003)^+$ ($\bar{D}^0D_s^{*+}$),
$Z_b(10610)^+$ ($\bar{B}^0B^{*+}$),
$Z_b(10650)^+$ ($\bar{B}^{*0}B^{*+}$); and some other states
are not close to such thresholds, such as the $Y(4360)$, $Y(4500)$,
$Y(4660)$, $Z_c(4430)^+$, and $Z_{cs}(4220)^+$.
These may suggest that we did observe the hadronic molecules
close to thresholds and we also observed hadronic states
with some other quark configurations like compact tetraquark
states and so on. 

It is expected that more results will be produced by the Belle II, 
BESIII, LHCb, and other experiments. Theoretical efforts are also
essential for understanding these new particles.

\section*{Acknowledgments}

I thank the organizers for the invitation and it is a pity I cannot
join the workshop in person due to the COVID-19 pandemic. 
This work is supported in part
by National Key Research and Development Program of China
(No.~2020YFA0406300), and National Natural Science Foundation
of China (NSFC, Nos.~11961141012, 11835012, and 11521505).

\end{document}